\documentclass[a4paper,11pt]{article}
\usepackage{pos}

\title{ Extension of Glauber-like model for Proton-Proton collisions using anisotropic and inhomogeneous density profile}
\author*[a]{Suman Deb}
\author[a]{Golam Sarwar}
\author[a]{Dhananjaya Thakur}
\author[b]{Pavish S.}
\author[a,c]{Raghunath Sahoo}
\author[d]{Jan-e Alam}

\affiliation[a]{Department of Physics, Indian Institute of Technology Indore,
  Simrol, Indore 453552, India}
  
  \affiliation[b]{Department of Astroparticle physics, Bergische Universit{\"a}t Wuppertal,
  Wuppertal, Germany}

\affiliation[c]{CERN, 
CH 1211, Geneva 23, Switzerland}

\affiliation[d]{Variable Energy Cyclotron Centre,
1/AF, Bidhan Nagar, Kolkata 700064, India}

\emailAdd{sumandeb0101@gmail.com}

\abstract{
Results from proton-proton ($pp$) collisions have routinely been used as a baseline to analyze and understand the production of QCD matter expected to be produced in nuclear collisions. 
But recent studies of small systems formed in $pp$ collisions at the LHC energies hint at the 
possibility of producing medium with collective behavior. 
Therefore, results from $pp$ collisions required more careful investigation to understand 
whether QCD matter is produced in high multiplicity $pp$ collisions.  With this motivation, 
the Glauber model traditionally used to study the heavy-ion collision dynamics at high 
energy is applied here to understand the dynamics of $pp$ collisions. 
We have used anisotropic and inhomogeneous quark/gluon-based proton density 
profile, a realistic picture obtained from deep inelastic scattering results
and this model explains the charged particle multiplicity distribution of $pp$ collisions at 
LHC energies very well. Collision geometric properties like impact parameter and mean number of 
binary collisions ($\langle N_{coll} \rangle$), mean number of participants 
($\langle N_{part} \rangle$) at different multiplicities are determined for $pp$ collisions. 
We further used these collision geometric properties to estimate average charged-particle 
pseudorapidity density ($\langle dN_{ch}/d\eta \rangle$) and found it to be comparable with 
the experimental results. Knowing $\langle N_{coll} \rangle$, we have obtained nuclear 
modification-like factor ($R_{pp}$) in $pp$ collisions which has not been done before 
to the best of our knowledge.}

\FullConference{%
 
 The Ninth Annual Conference on Large Hadron Collider Physics - LHCP2021\\
 7-12 June 2021\\
 Online
}

\begin{document}
\maketitle

\section{Introduction}

Quark-gluon plasma (QGP), a deconfined state of Quantum Chromodynamics (QCD) matter, is believed to be produced in heavy-ion collisions at ultra-relativistic energies. The results of such collisions are interpreted by considering proton-proton ($pp$) collisions as a baseline where QGP formation is unforeseen. However, recent results~\cite{ALICE:2017jyt,Alver:2010ck,Khachatryan:2010gv} show that high multiplicity $pp$ collisions behave like heavy-ion collisions. Therefore, it becomes crucial to examine the dynamics of the QCD medium formed in ultra-relativistic $pp$ collisions carefully. In this article, we present the results of  Glauber-like model calculations for $N_{coll}$(b), $N_{part}$(b) due to the quark and gluon-based proton density profile. In the case of a proton, several density profiles have been considered to estimate the initial conditions, most of them assume an azimuthally symmetric density profile, and those are mainly different in the phenomenological parametrization of radial variations~\cite{dEnterria:2010xip}. In this regard, we find in Ref.~\cite{Kubiczek:2015zha} to consider the azimuthally asymmetric and inhomogeneous density distribution of a proton . This is close to a realistic picture obtained by results of deep inelastic scattering revealing the structure of proton. We have used the proposed model~\cite{Kubiczek:2015zha} with the modifications explained in \cite{Deb:2019yzc}, to obtain the charged particle multiplicity distribution in $pp$ collisions at $\sqrt{s}$ = 7 TeV and compared it with the ALICE result. To understand the possibility of medium formation in high-multiplicity $pp$ collisions and having calculated $N_{coll}$, we have estimated a nuclear modification-like factor, $R_{pp}$, considering low multiplicity yields as the base.

\section{FORMALISM}

Here we consider a fluctuating proton orientation with three effective quarks and gluonic flux tubes connecting them. The density function of the resultant proton considered in this work, taken from Ref.~\cite{Kubiczek:2015zha}, given by
\begin{equation} 
\rho_{G-f} (\textbf{r} ; \textbf{r}_{1} ,\textbf{r}_{2} ,\textbf{r}_{3} )  = N_{g} \frac{1 - \kappa}{3} \sum\limits_{ k = 1}^{3} \rho_{q}(\textbf{r} - \textbf{r}_{k} ; r_{q} ) +  N_{g} \frac{\kappa}{3} \sum\limits_{ k = 1}^{3} \rho_{g}[\mathcal{R}^{-1}[\theta_{k} ,  \phi_{k} ](\textbf{r} -    \frac{\textbf{r}_{k}}{2} ; r_{q} , \frac{r_{k}}{2}\textbf]
\label{rho_together}
\end{equation} 
where $r_{q}$ is the radius of the quark,  $r_{s}$ and $r_{l}$ are the radius and the length of the gluon tube. The first term (second term) of the above equation represents the densities of quarks (gluons). Further, $\mathcal{R[\theta , \phi ]}$ transforms the vector (0,0,1) into $(cos\phi sin\theta , sin\phi cos\theta , cos\theta), $ and $\textbf{r}_{k} = r_{k}(cos\phi_{k} sin\theta_{k} , \\ sin\phi_{k} cos\theta_{k} , cos\theta_{k}) $ (where, k = 1,2 and 3) is the position vector of the $k^{th}$ effective quark. $N_{g}$ is the number of partons inside the proton. The free parameter $\kappa$ allows to control the percentage of gluon body content and here it is taken to be 0.5~\cite{Kubiczek:2015zha}.

The collision plane is taken to be in x-y. Thus the dependence along the z-axis is integrated out. 
The overlap function $T_{pp}(b)$ for projectile proton (\textbf{A}) and target proton (\textbf{B}) is defined as
\begin{equation}
T_{pp}(b) = \int \int T_{A}(x - \frac{b}{2},y)T_{B}(x + \frac{b}{2},y) dxdy
\label{overlap_definition}
\end{equation}  
It is to be noted here that $T_{pp}$ is a combination of four terms due to quark-quark, quark-gluon, gluon-quark, gluon-gluon interactions due to the design of the density function. We now define $N_{coll}$  and $N_{part}$ of partons  in the $pp$ collisions at a given impact parameter (b) considering Ref.~\cite{Roli:glauber} as follows: 
\begin{equation}
N_{coll}(b) = \sigma_{gg}T_{pp}(b), {\text{and}} \\~
N_{coll}(b) \propto N_{part}^{x}(b)
\label{n_coll}
\end{equation} 
where x is a parameter, taken to be 0.75 as $N_{coll}$ scales as $A^{4/3}$ for similar target and projectile nuclei with mass numbers $A$ for heavy ion collisions and are spherical in shape. In line with the previous studies~\cite{Kubiczek:2015zha,Glazek:2016vkl}, we fixed the effective partonic cross-section ($\sigma_{gg}$) = 4.3 $\pm$ 0.6 mb~\cite{Drescher:2007cd} with 
$N_{g}$= 10 partons to reproduce the experimental value of the inelastic $pp$ cross-section,
($\sigma_{pp}$) = 60 mb at $\sqrt{s}$ = 7 TeV. This gives the only non-trivial dependence of the model on the beam energy $\sqrt{s}$. 

\section{Results}
\subsection{\textbf{Charged particle multiplicity estimation}} 
\begin{figure}
\begin{center} 
  \includegraphics[scale=0.34]{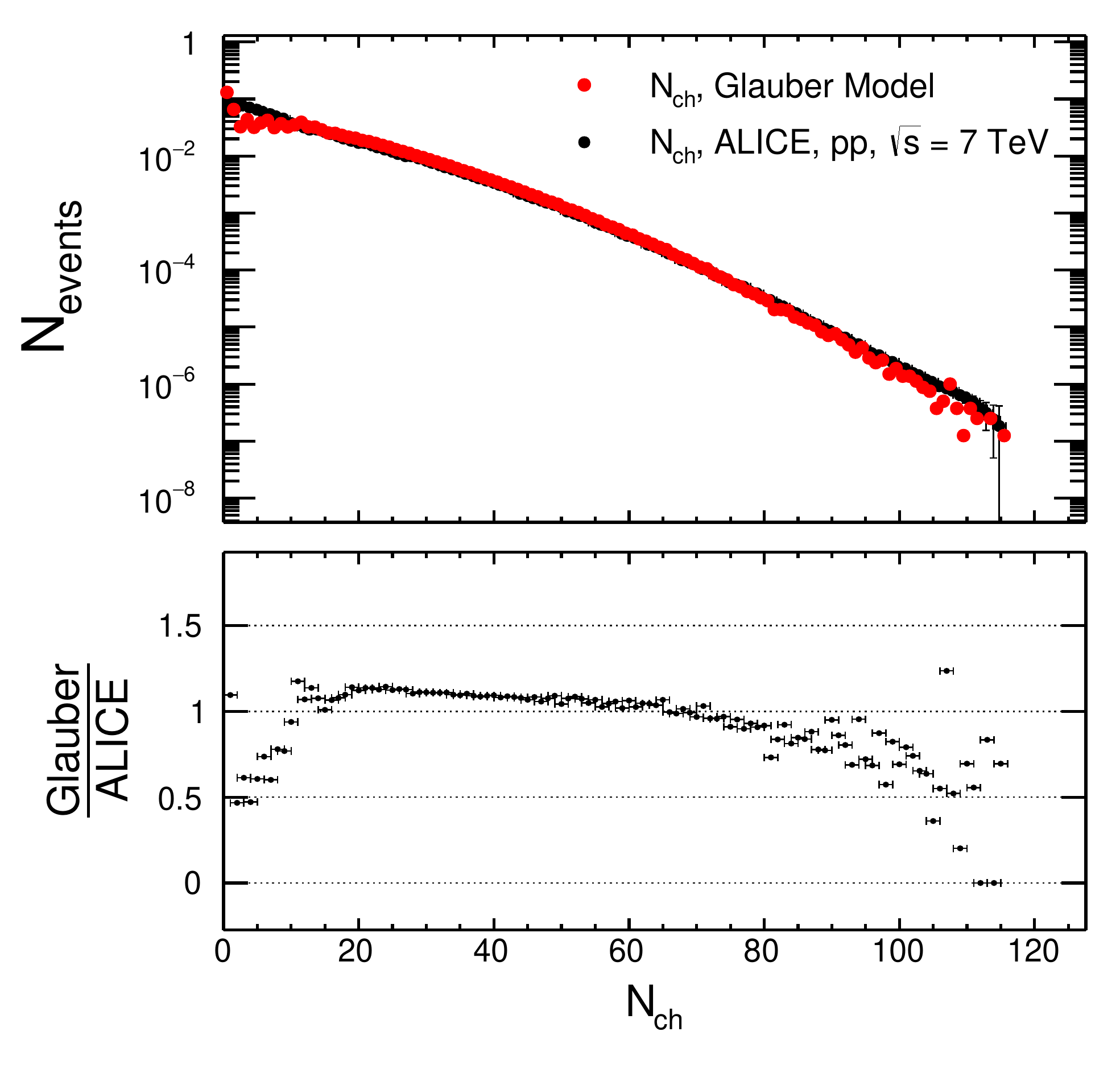}
  \caption{Upper panel: Comparison of the charged multiplicity distribution obtained from the present work with the ALICE experiment, for $pp$ collisions at $ \sqrt{s} =$ 7 TeV. Black dots represent ALICE data and red dots represent the present work. Lower panel: Ratio of this work to the ALICE experimental data \cite{Deb:2019yzc}.}
   \label{fig:alicedata_glouber}    
   \end{center} 
  \end{figure}

In this work, we have used the two-component model (commonly used in heavy-ion phenomenology) approach to the $pp$ collisions, replacing nucleons by partons (quarks and gluons) and the nucleon-nucleon inelastic cross-section by the effective partonic cross-section ($\sigma^{inel}_{gg}$). Thus, the number of ancestors (independently emitting sources of particles) as parametrized by a two-component model is given as 
\begin{equation}
N_{ancestors} = f N_{part} + (1-f) N_{coll}
\label{ancestor}
\end{equation} 
where $f$ is a free parameter.  As the negative binomial distribution (NBD) successfully reproduces the charged-particle distribution in $pp$ collisions~\cite{Alner:1985rj}, we use two-parameters NBD to calculate the probability of producing $n$ particles per ancestor where $\bar{n}$ is the average multiplicity, and $k$ characterizes the width of the distribution. The best agreement for the $N_{ch}$ distribution obtained by our model with experimental data is found for $f=$ 0.85, $\bar{n} = 8$ and $k = 0.13$.  From Fig.~\ref{fig:alicedata_glouber}, it can be seen that our model well describes the data in the mid multiplicity region ( $15 < N_{ch} < 90$ ), with a 5-10 \% discrepancy. However, towards the low and high multiplicity, it is unable to reproduce the experimental measurement. The inability of the model to explain the extreme low and high multiplicity region might be due to a lack of statistics. 

  \subsection{\textbf{The ratio, $R_{pp}$ for high to low multiplicity events}}
\begin{figure}
  \includegraphics[scale=0.305]{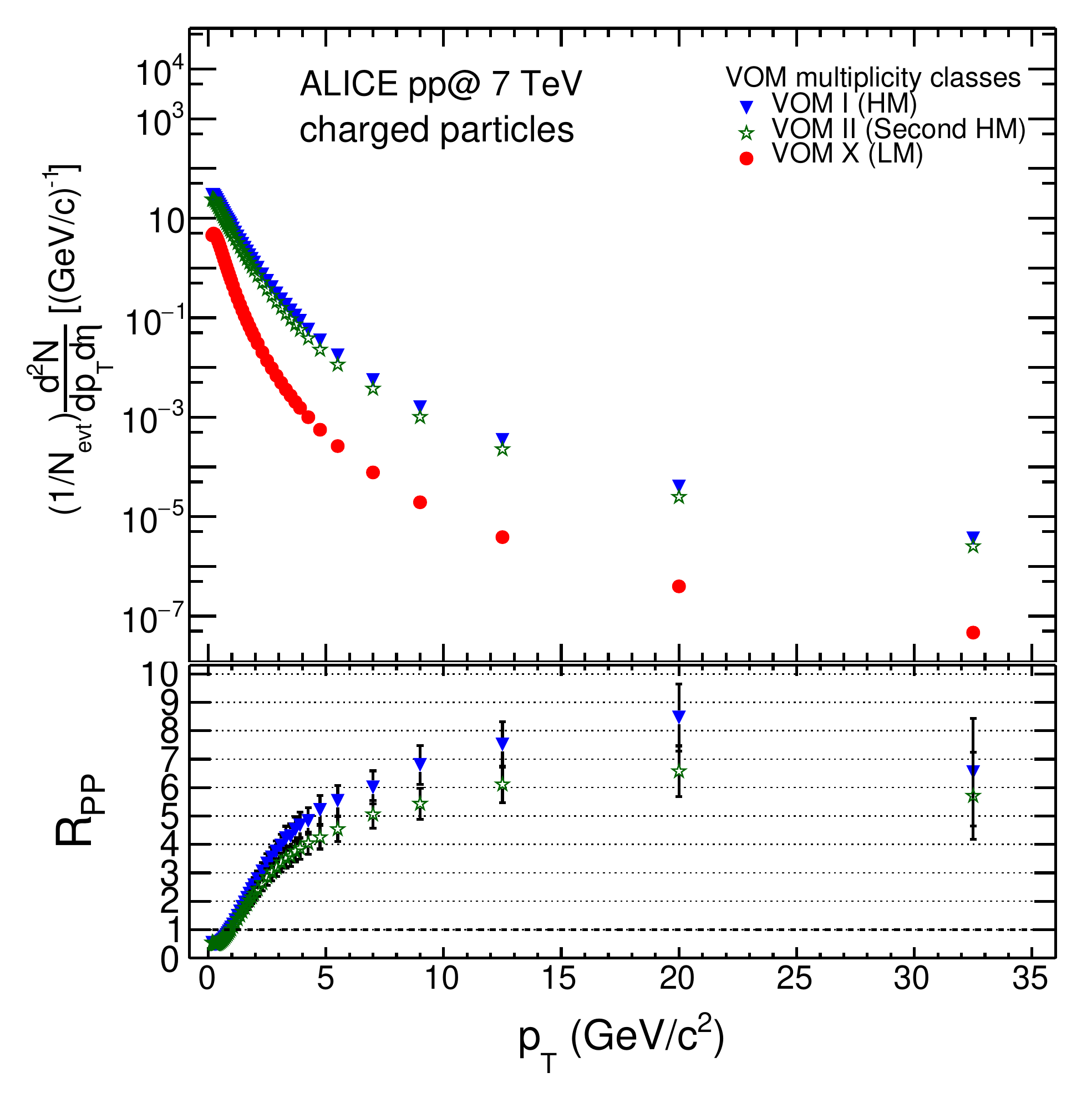}
    \includegraphics[scale=0.405]{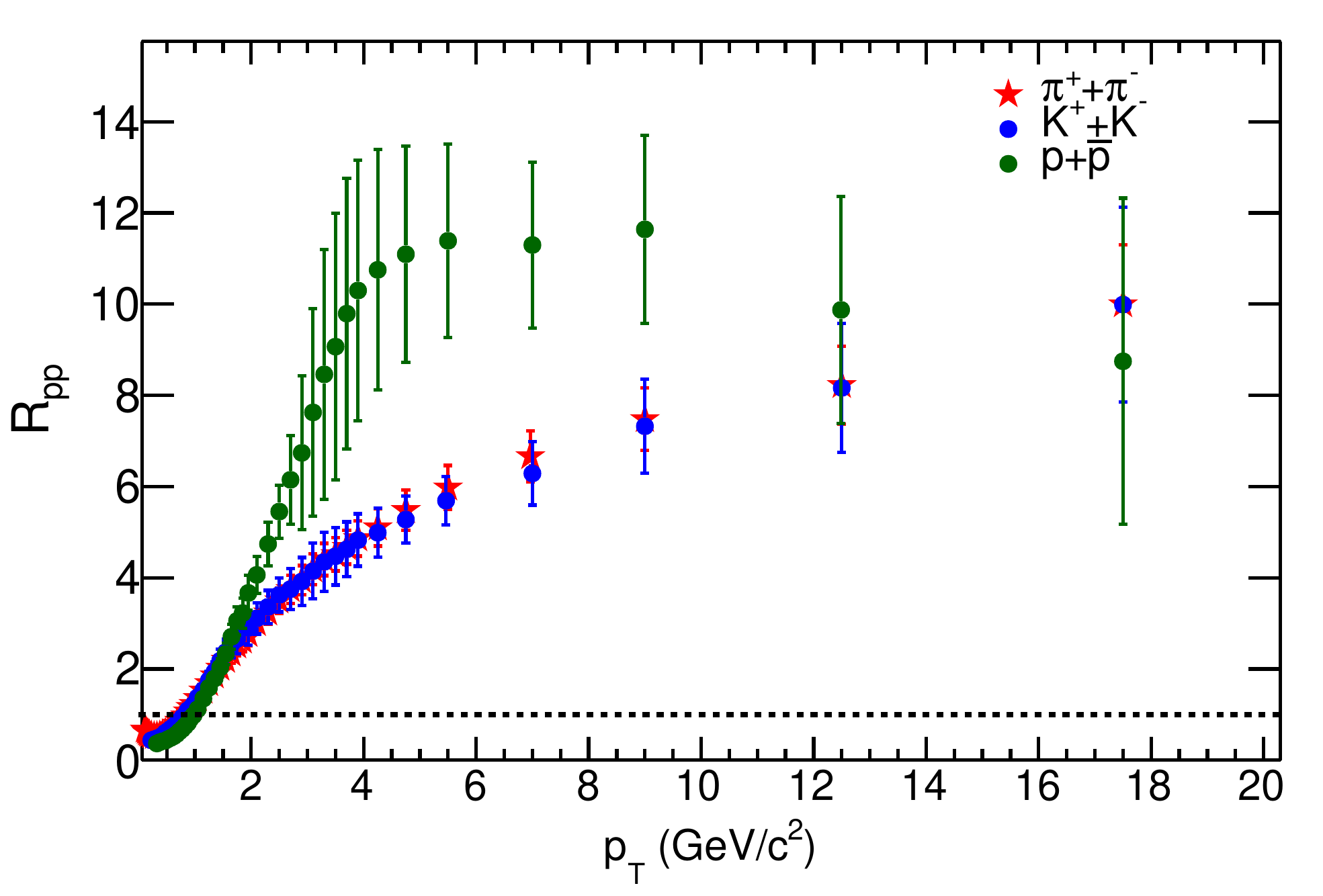}
  \caption{\cite{Deb:2019yzc} Upper panel (left): Transverse momentum spectra of charged particle in $pp$ collisions at $\sqrt{s}$ = 7  TeV~\cite{Acharya:2018orn} for VOM multiplicity classes, viz., highest (HM), second highest (second HM) and lowest multiplicity (LM) class. Lower Panel (left): $R_{pp}$  obtained from the ratio of differential yield at HM and second HM with low multiplicity class scaled by $\langle N_{coll} \rangle$. Right: Nuclear modification-like factor obtained from Eq.~\ref{Rpp} for pion, Kaon and proton in $pp$ collisions at $\sqrt{s}$ = 7  TeV~\cite{Acharya:2018orn}.  }  
 \label{fRpp}     
  \end{figure}

To understand the possibility of the formation of a medium in high-multiplicity events in $pp$ collisions, we define a quantity, $R_{pp}$ (Fig.~\ref{fRpp}) as: 
\begin{equation}
 R_{pp}(p_{T}) = \frac{d^{2}N/dp_{T}d\eta|^{HM}}{d^{2}N/dp_{T}d\eta|^{LM}} \times \frac{\langle N_{coll}^{LM} \rangle}{\langle N_{coll}^{HM} \rangle}
\label{Rpp}
\end{equation} 
which is similar to the nuclear modification factor ($R_{AA}$) in heavy-ion collisions. Here, $d^{2}N/d\eta dp_{T}|^{HM}$ ($d^{2}N/d\eta dp_{T}|^{LM}$), $\langle N_{coll}^{LM} \rangle$ ($\langle N_{coll}^{HM} \rangle$) are charged particle yield in high (low)-multiplicity, mean number of binary collisions in low (high) multiplicity for $pp$ collisions at $\sqrt{s}$ = 7  TeV~\cite{Acharya:2018orn}. 

\section{Summary}

In this contribution, we have used an anisotropic and inhomogeneous proton density profile as an initial condition for $pp$  collisions. We have obtained the charged-particle multiplicity distribution for $pp$ at $\sqrt{s}$ = 7 TeV and compared the result with ALICE. It is found that the proposed model can well reproduce the multiplicity distribution originating from $pp$ events with the free parameter $f=0.85$. We have also estimated a nuclear modification-like factor ($R_{pp}$) for $pp$ collisions and found that the factor is less than $1$  for $p_{\it T}<1$ GeV, and beyond this, the factor is greater than unity. This behavior at higher $p_{\it T}$ may be due to non-collective flow effects, which need further investigation. The present work is important as a calculation based on the Glauber-like model is proposed for the $pp$ system with modifications to other works like~\cite{Kubiczek:2015zha,dEnterria:2010xip}. This work has appeared as a regular publication as Ref.~\cite{Deb:2019yzc}.
 
\end{document}